\documentclass[preprint,showpacs,preprintnumbers,groupedaddress,amsmath,amssymb,aps,pra]{revtex4}

\usepackage{graphicx}
\usepackage{bm}

\begin{document}

\title{Two-pulse interference and superluminality}

\author{Bruno Macke} 
\author{Bernard S\'{e}gard} 
\email{bernard.segard@univ-lille1.fr}
\affiliation{Laboratoire de Physique des Lasers, Atomes et Mol\'{e}cules (PhLAM), 
Centre d'Etudes et de Recherches Lasers et Applications (CERLA), 
Universit\'{e} de Lille I, 59655 Villeneuve d'Ascq, France}

\date{\today}

\begin{abstract}

We examine how the interference of a coherent light-pulse with its slightly \emph{time-delayed} copy may generate a pulse nearly identical to the original one and \emph{ahead} of it. The simplicity of this 2-pulse system enabled us to obtain exact analytic expressions of the pulse distortion, valid for a wide class of pulse shapes. Explicit results are given for the pulses usually considered (gaussian, hyperbolic secant) but also for more realistic pulses of strictly limited duration. We finally show that the \emph{efficiency} of the 2-pulse system is comparable to that of the other superluminal systems, at least for the pulse advancements actually demonstrated in the optical experiments.

\end{abstract}
\maketitle

Although unambiguous demonstrations of the phenomenon have been achieved in the 1980's \cite{ref1,ref2}, the now called superluminal propagation of light pulses (fast light) still attracts considerable interest. For reviews see, e.g., \cite{ref3,ref4,ref5}. In an ideal experiment, the pulse having covered some distance $L$ in a linear medium with anomalous dispersion leaves it with almost no change in shape, in significant advance of a pulse which would have covered the same distance $L$ at the velocity $c$ of light in vacuum. As a matter of fact, the field transmitted at some time  $t$ is not the consequence of the signal entering the medium at the same time $t$ but that of all the signals anterior to $t$ by more than $L/c$  and this puzzling phenomenon is not at odds with the relativistic causality.  It is now well recognised that the superluminal propagation originates in a predominantly destructive interference between the different frequency-components of the pulse, all propagating with velocities smaller than $c$ (see, e.g.,\cite{ref6}). A similar mechanism occurs in systems involving a non-dispersive medium when several modes or paths of light-transmission, with different propagation times, are possible. The interference of the different transmitted pulses may then generate an output pulse ahead of the fastest one among them and even in absolute advance of the input pulse \cite{ref7,ref8}. This explains in particular the superluminal transmission of optical pulses through a multilayer structure \cite{ref9,ref10} though, due the high multiplicity of possible paths, the phenomenon is usually analysed in terms of evanescent waves (optical tunnelling). A simpler and more direct demonstration of multiple-pulse interference is provided by the recent experiment involving a fibre taper coupled with a microsphere \cite{ref11}. The output pulse then results from the interference of the pulses transmitted by the taper after having achieved $n$ roundtrips ($0\leq n <\infty$ ) inside the sphere (whispering-gallery mode). In order to examine how a time-delay can originate a time-advancement, we examine in this paper the still simpler situation in which only two pulses interfere (2-pulse systems). This occurs in particular in birefringent media where the light can propagate at two different velocities. Superluminality has then been inferred from phase measurements by Solly \emph{et al}. \cite{ref12} and directly demonstrated by Brunner \emph{et al}. \cite{ref13}. 

\begin{figure}[ht]
\begin{center}
    \includegraphics[angle=0,width=8cm]{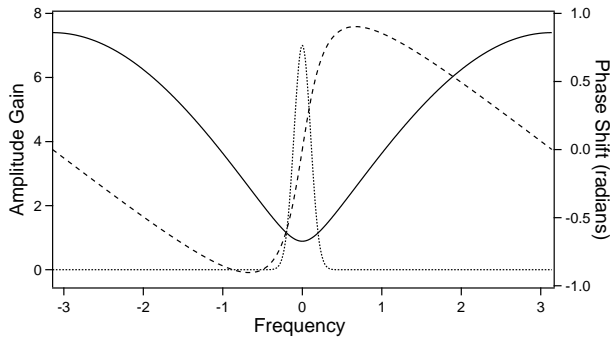}
    \caption{ Amplitude-gain $\left|H(\Omega)\right|$ (full line) and phase shift $\varphi(\Omega)$ (dashed line) of the 2-pulse system as a function of the detuning $\Omega=\omega-\omega_{0}$ , for $M=7.4$ and $m=0.89$. The frequency unit is $1/\tau$. The spectrum of a gaussian pulse of duration $\tau_{p}=13\tau$ is given for reference (dotted line). \label{fig1}}
\end{center}
\end{figure}

The causality principle fixes a strong link between the transmission-delay of a linear system and its gain profile \cite{ref14,ref15,ref16,ref5}. The largest superluminal effects are generally expected when the frequency $\omega_{0}$  of the optical carrier coincides with a well-marked minimum of gain. In the present case this is actually achieved when the 2-pulse interference is predominantly destructive, as indicated before. The relation between the slowly-varying envelopes of the incident and output pulses $e_{in}(t)$  and $e_{out}(t)$  reads then $e_{out}(t)=A e_{in}(t)-B e_{in}(t-\tau)$  with $A>B>0$ and $\tau>0$. The coefficient $A$ ($B$) is the amplitude-gain for the fast (slow) pulse and $\tau$ is the delay of the slowest pulse with respect to the fastest one ($t$ is thus a local time). Finally the 2-pulse system can be characterised by its impulse response $h(t)=A\delta(t)-B\delta(t-\tau)$ or, in the frequency domain, by its transfer function $H(\Omega)=A-B\exp(-i\Omega\tau)$, where $\Omega$ is the deviation of the current optical frequency $\omega$ from $\omega_{0}$. Our definitions and sign conventions are those used in signal theory \cite{ref17}. The phase shift $\varphi(\Omega)=\arg\left[H(\Omega)\right]$ and the amplitude-gain $\left|H(\Omega)\right|$  are respectively odd and even periodic functions of $\Omega$ , the latter oscillating between $m=A-B$ (its value for $\Omega=0$) and $M=A+B$ (constructive interference). Fig.\ref{fig1} shows the frequency dependence of $\left|H(\Omega)\right|$ and of $\varphi(\Omega)$ in a case taken as reference in the following.

Some light on the superluminality mechanism is obtained by doing a Taylor series expansion of $e_{out}(t)$. At the first order in $\tau$, we get		
\begin{equation}
e_{out}(t)\approx (A-B)e_{in}(t)+(-B)(-\tau)\frac{de_{in}(t)}{dt} \approx m\left[e_{in}(t)+a_{1}\frac{de_{in}(t)}{dt}\right]	\label{eq1}
\end{equation}

with $a_{1}=\left[\frac{M}{m}-1\right]\frac{\tau}{2}$. This leads at the first order in $a_{1}$ to $e_{out}(t)\approx m e_{in}(t+a_{1})$ and $H(\Omega)\approx m\exp(i\Omega a_{1})$.  Despite their apparent roughness, the previous approximations are exactly those currently used to define the group velocity. $a_{1}$ is nothing but the group time-advancement $d\varphi/d\Omega |_{\Omega=0}$  and is actually positive since the negative signs associated with the time-delay $\tau$ of the slow pulse and with the destructive interference compensate one another (see the intermediate form in Eq.\ref{eq1}).  At this order of approximation, the output pulse has the same shape that the input pulse, is $m$ times larger and is ahead of the fastest transmitted pulse by $a_{1}$. The approximation will be good if $m\exp(i\Omega a_{1})$ is itself a good approximation of the exact transfer function on the \emph{entirety} of the pulse spectrum. This requires (at least) the time-delay $\tau$ to be small compared to the duration $\tau_{p}$ of the pulse (hereafter the full width at half maximum of its intensity profile).  The examination of the expression of $a_{1}$ then shows that group advancements comparable to the pulse duration can be attained only if the maximum gain $M$ is large compared to the gain $m$ at the frequency $\omega_{0}$ of the optical carrier. This property is common to all the superluminal systems \cite{ref18}. In purely passive (not amplifying) optical systems $M\leq1$ and the condition $M/m>>1$ is only met when $m<<1$. A large post-amplification (often not explicitly mentioned) is then required to normalise the amplitude of the output pulse. Conversely some post-attenuation may be required in active systems to obtain $m\approx1$ \cite{ref19}. In both cases the device achieving the normalisation of the output pulse should be considered as a part of the superluminal system and the gains considered in the following (as well as in Fig.\ref{fig1}) are those of the overall system. Note however that the double condition  $M>>m\approx1$ can be met in all-optical systems, for example in an amplifying medium with a dip in its gain-profile, as proposed in \cite{ref16} and demonstrated in \cite{ref20}.

Superluminal effects are always accompanied by some pulse distortion. The distortion is negligible only when the spectrum (duration) of the input pulse is very narrow (long) but the fractional advancement is then also negligible. Although this point is often not addressed, \emph{the real challenge is to attain significant fractional advancements with moderate pulse distortion}. The latter is conveniently characterised by the root-mean-square (\emph{rms}) deviation $D$ of $e_{out}(t)$ from the ideal advanced form $e_{in}(t+a)$ , where $a$ is the aimed advancement. By normalising the energy of the input pulse ($\int_{-\infty}^{+\infty}\left|e_{in}(t)\right|^{2}dt=1$), $D$ is reduced to $D=\left(\int_{-\infty}^{+\infty}\left|e_{out}(t)-e_{in}(t+a)\right|^{2}dt \right)^{1/2}$.  Minimising the distortion is in general a formidable problem without analytical solution \cite{ref18}. The 2-pulse system considered here is probably the only system for which exact analytical results can be obtained. For sake of simplicity, we assume that $e_{in}(t)$ is real (no chirping). $e_{out}(t)$ is then also real and $D$ depends on the pulse shape only via $p(\tau)$, $p(a)$ and $p(a+\tau)$, where $p(t)=\int_{-\infty}^{+\infty} e_{in}(t+\theta)e_{in}(\theta)d\theta$ is the autocorrelation of  $e_{in}(t)$ \cite{ref17}.  Recall that $p(t)$ is an even function of $t$, has its maximum (equal to 1) for $t=0$, and is the inverse Fourier transform of the energy spectrum of $e_{in}(t)$. For given delay $\tau$ and advancement $a$, simple calculations show that $D$ attains a minimum
\begin{equation}
D_{min}(a,\tau)=  \sqrt{1-\frac{p^{2}(a)+p^{2}(a+\tau)-2p(\tau)p(a)p(a+\tau)}{1-p^{2}(\tau)}} \label{eq2}
\end{equation}
when $M=\frac{p(a)-p(a+\tau)}{1-p(\tau)}$ and $m=\frac{p(a)+p(a+\tau)}{1+p(\tau)}$. The distortion only cancels in the trivial case $a=0$. We have then $M=m=1$ , that is $A=1$ and $B=0$ (no interference). For $a>0$, the minimum distortion is attained for $m<1$ and the group advancement $a_{1}$ generally differs both from $a$ and from the advancement $a_{m}$ of the pulse maximum. It keeps however a physical meaning. A simple application of the so-called moment theorem in signal analysis \cite{ref17} indeed shows that $a_{1}$ is the advancement of the centre-of-gravity of the pulse envelope, as large as the distortion may be \cite{ref16}. If $e_{in}(t)$ is continuous, the derivatives $p^{(n)}(t)=d^{n}p/dt^{n}$ of $p(t)$ are continuous up to $n=2$ and the distortion has an absolute minimum $D_{amin}$ when $\tau\rightarrow0$. It reads 
\begin{equation}
D_{amin}(a)=\sqrt{1-p^{2}(a)+\frac{\left[p^{(1)}(a)\right]^{2}}{p^{(2)}(0)}}	\label{3}
\end{equation}
where $p^{(2)}(0)<0$. At this limit, $m=p(a)$, $a_{1}=\frac{p^{(1)}(a)}{p(a)p^{(2)}(0)}$ and $M=\frac{2p^{(1)}(a)}{\tau p^{(2)}(0)}$.  The latter expression shows that, for a given advancement, the lowest distortion is attained when the maximum gain $M$ of the system tends to infinity. When $M$ is large but finite, we easily get
\begin{equation}
D^{2}_{min}(a,M)\approx	D^{2}_{amin}(a)  +\frac{2\left[p^{(1)}(a)\right]^{2}}{M p^{(2)}(0)}\left[\frac{p^{(2)}(a)}{ p^{(2)}(0)}-p(a)\right] \label{eq4}
\end{equation}
that is $D_{min}(a,M)/D_{amin}(a)=1+O(1/M)$. The asymptotic limit $D_{amin}$ can thus be approached for moderate values of $M$.  The above results are valid under the unique assumption that the pulse envelope $e_{in}(t)$ is continuous. Since discontinuities in $de_{in}/dt$ may originate strong transients in the output pulse (see, e.g., Fig.3 in \cite{ref21}), we will now consider the usual case of bell-shaped envelopes, the derivative of which is continuous. The autocorrelation $p(t)$ is then also bell-shaped, its derivatives $p^{(n)}(t)$ are continuous up to $n=4$ and it is possible to obtain approximate expressions of $D_{amin}$, $m$ and $a_{1}/a$ valid at the 2nd order in $a$. They read  $D_{amin}(a)\approx\frac{a^{2}}{2}\sqrt{p^{(4)}(0)-\left[p^{(2)}(0)\right]^{2}}$, $m\approx 1 +p^{(2)}(0)\frac{ a^{2}}{2}$ and $\frac{a_{1}}{a}\approx 1+\frac{a^{2}}{2} \left[\frac{p^{(4)}(0)}{3p^{(2)}(0)}-p^{(2)}(0) \right]$. At this level of approximation, the distortion is mainly due to the frequency-dependence of the gain (Fig.\ref{fig1}), which originates a broadening of the pulse spectrum and thus a narrowing of the pulse itself \cite{ref16}. Due to the pulse-area-conservation theorem \cite{ref22}, this narrowing is accompanied by a pulse magnification and this explains why the \emph{rms} distortion is minimised when the gain $m$ at $\Omega=0$ (that is at the centre of the pulse spectrum) is slightly smaller than unity. 

The previous results apply to a wide class of pulse shapes. We first consider the gaussian pulses, the most popular ones in the literature on superluminality. We write their envelope under the form $e_{in}(t)\propto \exp(-t^{2}/2)$, such that $\tau_{p}=2\sqrt{\ln2}\approx 1.67$ and $p(t)=\exp(-t^{2}/4)$. We then get $D_{amin}(a)=\sqrt{1+\left(1+a^{2}/2\right)\exp\left(-a^{2}/2\right)}$ and and $a_{1}=a$. Note however that the common value of $a_{1}$ and $a$ differs from the advancement $a_{m}$ of the pulse maximum. The latter is easily determined from the expression of $e_{out}(t)$ and reads $a_{m}=\left(\sqrt{1+4a_{1}^{2}}-1\right)/2a_{1}$. At the 2nd order in $a$, we finally get $D_{amin}(a)\approx a^{2}/2\sqrt{2}\approx 0.98\left(a/\tau_{p}\right)^{2}$  with $m\approx 1-a^{2}/4$ and $a_{m}/a \approx 1-a^{2}$. 

As a second example we consider the classical hyperbolic-secant pulses (sech-pulses) $e_{in}(t)\propto 1/\cosh t$, such that $\tau_{p}=2\arg\cosh(\sqrt{2})\approx 1.76$  and $p(t)=t/\sinh t$. We then get $D_{amin}(a)=\sqrt{1-\left(\frac{a}{\sinh a}\right)^{2}\left[3\left(\coth a-\frac{1}{a}\right)^{2}+1\right]}$  and $a_{1}=3\left(\coth a-1/a\right)\neq a$ , that is, in the 2nd order approximation, $D_{amin}\approx 2a^{2}/3\sqrt{5} \approx 0.93\left(a/\tau_{p}\right)^{2}$ , with $m\approx 1-a^{2}/6$ and $a_{m}/a \approx 1-a^{2}/15$. 

Contrary to the analytical pulses above considered, the pulses actually used in the experiments have obviously a strictly finite duration and their envelope has thus at least two points of non-analyticity. There is then an unavoidable distortion due to the fact that, in our local time picture, the output signal cannot anticipate the beginning and the end of the input signal. Moreover, when $M$ is very large, these points of non-analyticity may generate strong transients in the output pulse \cite{ref21}, reinforcing the distortion. All that occurs even if the envelope of the input pulse is quasi-gaussian. In the sequel we will take as reference the "cos-pulses" of envelope $e_{in}(t)\propto (1+\cos t)$,  for $\left|t\right|<\pi$ and $e_{in}(t)=0$ elsewhere (total width $2\pi$). They are such that $\tau_{p}=2\arccos(\sqrt{2}-1)\approx 2.29$, $p(t)=\frac{2+\cos\left|t\right|}{3}\left(1-\frac{\left|t\right|}{2\pi}\right)+\frac{\sin\left|t\right|}{2\pi}$  for $\left|t\right|<2\pi$ and $p(t)=0$  elsewhere. We now get $D_{amin}(a)=\sqrt{1-p^{2}(a)-3f^{2}(a)}$, $a_{1}/a=3f(a)/ap(a)\neq 1$ and $a_{m}=\arctan(a_{1})$, with $f(a)=\left(1-\frac{a}{2\pi}\right)\frac{\sin a}{3}+\frac{1-\cos a}{3\pi}$ . At the 2nd order in $a$, these expressions read $D_{amin}\approx a^{2}/3\sqrt{2} \approx 1.23\left(a/\tau_{p}\right)^{2}$, $a_{1}/a \approx 1$ and $a_{m}/a \approx 1-a^{2}/3$, with $m\approx 1-a^{2}/6$ . We observe that, for moderate fractional advancements $a/\tau_{p}$, the distortion of the cos-pulses in the 2-pulse system is not much larger than that of the standard gaussian pulses. Note that, in both cases, the advancement $a_{m}$ is smaller than the group advancement.
 
\begin{figure}[ht]
\begin{center}
    \includegraphics[angle=0,width=8cm]{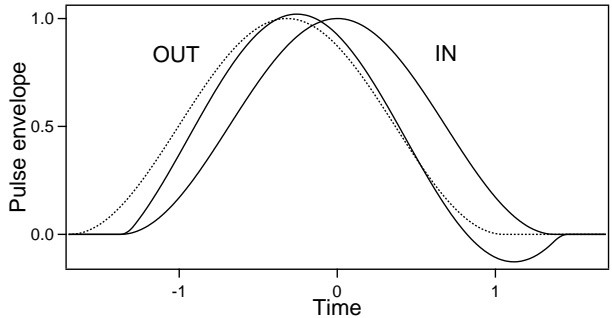}
    \caption{ Envelopes $e_{in}(t)$ of the input cos-pulse and $e_{out}(t)$ of the output pulse in the 2-pulse system for $a/\tau_{p}=0.32$ and $M=7.4$ (full line). The ideally advanced envelope $e_{in}(t+a)$ is given for comparison (dotted line). The time unit is the duration $\tau_{p}$ of the input pulse. \label{fig2}}
\end{center}
\end{figure}

\begin{figure}[ht]
\begin{center}
    \includegraphics[angle=0,width=8cm]{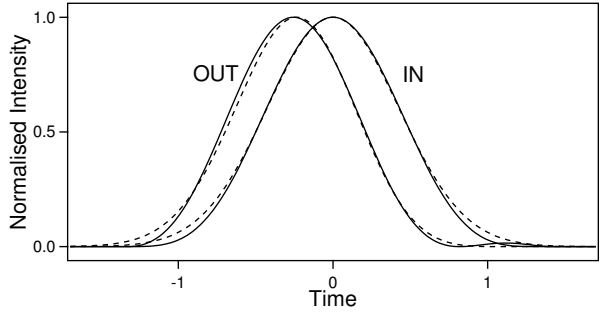}
    \caption{Normalised intensity-profiles corresponding to the pulse envelopes $e_{in}(t)$ and $e_{out}(t)$ of Fig.\ref{fig2} (full line). The profiles obtained with a gaussian pulse of same duration $\tau_{p}$ are given for reference (dashed line). \label{fig3}}
\end{center}
\end{figure}

As indicated before (see Eq.\ref{eq4}), the asymptotic limit $D_{amin}$ can be approached for reasonable values of the maximum amplitude-gain $M$. For further comparisons, we consider the case $a/\tau_{p}=0.32$ and $M=7.4$. The corresponding gain and phase profiles are those of Fig.\ref{fig1}. The distortion remains moderate ($14\% $), exceeding its absolute minimum $D_{amin}$ only by a factor $1.25$. We compare Fig.\ref{fig2} the envelope $e_{out}(t)$ to its ideal form $e_{in}(t+a)$.
As expected, one can see large deviations at the beginning and the end of the input pulse. Occurring in the pulse pedestal, these features are obviously much less apparent on the corresponding intensity-profile (Fig.\ref{fig3}), which is the profile usually observed in the optical experiments. Apart from a slight overshot at the end, the profile of the output pulse is comparable to that obtained with an input gaussian pulse of same duration. In both cases, the main distortion consists in a narrowing of the pulse, resulting in different advancements on its fall and on its rise.

\begin{figure}[ht]
\begin{center}
    \includegraphics[angle=0,width=8cm]{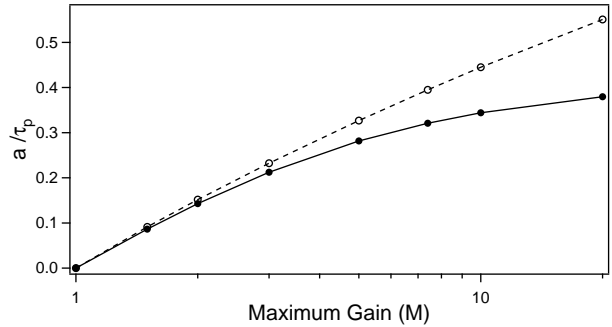}
    \caption{Fractional advancement $a/\tau_{p}$ attainable with an acceptable distortion (see text) as a function of the maximum gain $M$ (including the gain of the normalisation device). The input pulse is a cos-pulse. The full (dotted) line and the full (open) circles relate to the 2-pulse (single-dip) system. \label{fig4}}
\end{center}
\end{figure}

\begin{figure}[ht]
\begin{center}
    \includegraphics[angle=0,width=8cm]{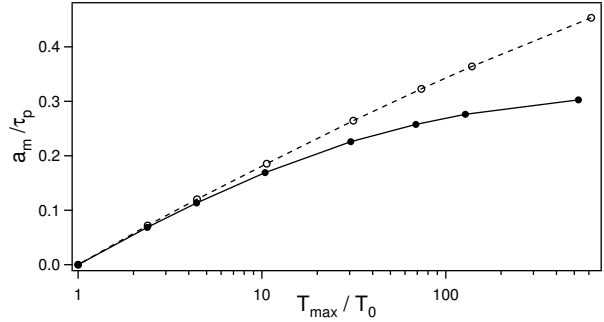}
    \caption{ Fractional advancement  $a_{m}/\tau_{p}$ of the pulse maximum as a function of the ratio $T_{max}/T_{0}$ of the maximum intensity-transmission of the optical system over its value at the centre of the pulse spectrum. The input pulse is a cos-pulse. The full (dotted) line and the full (open) circles relate to the 2-pulse (single-dip) system. \label{fig5}}
\end{center}
\end{figure}

In the abundant literature on superluminality, one finds few \emph{direct} demonstrations of \emph{significant} pulse-advancements (say $a>\tau_{p}/10$) with \emph{moderate} distortion \cite{ref2,ref11,ref19,ref23,ref24,ref25,ref26,ref27,ref28,ref29}. Indeed, problems of noise, of instability and of hypersensitivity to small defects in the input pulse severely limit the maximum gain $M$ which can be actually used in the experiments (keep in mind that we include in the system gain that of the device eventually used to normalise the output pulse, so that $m \lessapprox 1$ in the moderate distortion limit). In our experiment at a wavelength $\lambda \approx 3$ mm \cite{ref2}, we succeeded in using $M \approx 160$. In the optical domain, quantum effects (amplified spontaneous emission, detection noise, etc…) \cite{ref30} further reduce the gains actually usable and, to our knowledge, all the experiments have been achieved with $M<10$ (amplitude gain). Due to these restrictions on the maximum gain, the \emph{efficiency} of a superluminal system may be defined as its ability, $M$ being given, to attain a fractional advancement $a/\tau_{p}$ as large as possible with a moderate distortion. The acceptable distortion obviously depends on the advancement, very small when the fractional advancement is itself very small \cite{ref31} and the larger the larger $a/\tau_{p}$ is. In the following we retain as acceptable the distortion observed in our reference case (Fig.\ref{fig2} and Fig.\ref{fig3}), namely $D \approx 0.45 a/\tau_{p}$ . Fig.\ref{fig4} shows the fractional advancements $a/\tau_{p}$ attained in this way for various values of $M$ ranging from $1$ (no advancement) to $20$.
The advancement actually measured in the experiments is the advancement $a_{m}$ of the pulse maximum ($a_{m}<a$) and the most accessible parameter is the contrast $T_{max}/T_{0}$ between the maximum intensity-transmission of the optical device and its value at $\omega_{0}$. Since $T_{max}\propto M^{2}$ and $T_{0}\propto m^{2}$, we obviously have  $T_{max}/T_{0}=M^{2}/m^{2}\gtrapprox M^{2}$. The dependence of $a_{m}/\tau_{p}$ on $T_{max}/T_{0}$ is shown Fig.\ref{fig5}.

For comparison consider now the superluminal systems having demonstrated the most striking fractional advancements \cite{ref2,ref11,ref19,ref23,ref24,ref25,ref26,ref27,ref28,ref29}. Most involve an homogeneous medium, the transmission curve of which has a dip at the frequency $\omega_{0}$. The dip may be natural \cite{ref2,ref24} but is often induced by an additional light field \cite{ref25,ref26,ref27,ref28,ref29} (which unfortunately may also induce extra noise). Using our general conventions, $H(\Omega)=M \exp\left(-\frac{Z}{1+i \Omega/\gamma}\right)$ is generally a good approximation of the transfer function of these single-dip systems. In this expression $Z$ is the field optical-thickness associated with the dip and $\gamma$ is the width of the corresponding line, assimilated to a Lorentzian \cite{ref16}. The group advancement, the amplitude gain at $\Omega =0$ and the intensity-contrast $T_{max}/T_{0}$ respectively read $a_{1}=Z/\gamma$, $m=M\exp(-Z)$ and $T_{max}/T_{0}=\exp(2Z)$. The distortion can easily be expressed in the frequency domain by means of the Parseval's theorem \cite{ref17}. We get [16] $D=\sqrt{\int_{-\infty}^{\infty}\left|H(\Omega)-\exp(i\Omega a)\right|^{2}\left|E_{in}(\Omega)\right|^{2} d\Omega/2\pi}$ where $E_{in}(\Omega)=\int_{-\infty}^{\infty}e_{in}(t)\exp(-i\Omega t)dt$ is the Fourier transform of $e_{in}(t)$. $M$ and $a$ being fixed, the parameters $Z$ and $\gamma$ are adjusted in order to minimise the distortion. This is efficiently achieved by means of a genetic algorithm \cite{ref32}. Rapid convergence is obtained by exploring values of $Z$ and $\gamma$ close to those giving $m=1$ and $a_{1}=a$, that is $Z=\ln M$ and $\gamma=\ln M/a$. For each value of $M$, the fractional advancement $a/\tau_{p}$ is calculated by a dichotomy method: (i) Choose $0< a/\tau_{p}<1$ (ii) Minimise $D$ by the genetic algorithm (iii) If $D_{min}$ is too large (small), decrease (increase) $a/\tau_{p}$. Go to (i). We stop the calculation when $D_{min}$ equals the acceptable distortion within $0.1\%$ and we determine the corresponding advancement $a_{m}$ of the pulse maximum. Fig.\ref{fig4} (Fig.\ref{fig5}) shows $a/\tau_{p}$ ($a_{m}/\tau_{p}$) as a function of $M$ ($T_{max}/T_{0}$) \cite{ref33}. Comparing the advancements $a_{m}$ attained with the 2-pulse system and with the single-dip system, we see that the former is only $22\%$ smaller than the latter for the largest ratio $T_{max}/T_{0}$ actually involved in an optical experiment ($T_{max}/T_{0}\approx 100$). In most experiments $T_{max}/T_{0}<10$ and the two advancements are nearly equal. Conversely the single-dip system would be much more efficient if ratios $T_{max}/T_{0}>> 100$ could be used. Quite similar results are obtained by considering the other superluminal systems having demonstrated significant fractional advancements, in particular the periodic fibre Bragg gratings \cite{ref23}, the taper coupled with a microsphere \cite{ref11} and the doublet of gain lines \cite{ref19}. A better efficiency is expected from a doublet of absorption lines \cite{ref16,ref34} but, to our knowledge, this system has not been actually used.

To summarise, we have studied the superluminal system involving the predominantly destructive interference of a pulse with a time-delayed copy. Due to the simplicity of this system, we have succeeded in obtaining exact analytical expressions of the distortion. Finally we have shown that, despite its simplicity, this system permit to observe superluminal effects comparable to all those which have been actually demonstrated in optics.

Laboratoire PhLAM is Unit\'{e} Mixte de Recherche de l'Universit\'{e} de Lille I et du CNRS (UMR 8523).
 CERLA is F\'{e}d\'{e}ration de Recherche du CNRS
(FR 2416).


\begin{thebibliography}{10}
\bibitem{ref30} For a discussion of the quantum effects (not considered in the present paper), see Ref.\cite{ref5} and references therein.
\bibitem{ref33} Insofar as the prevailing distortion originates in the 2nd order variations of the gain, $D_{min}\propto a^{2}/\ln M$   for a gaussian pulse \cite{ref16}, so that $a\propto\left(D_{min}/a\right)\ln M$ . The nearly linear curve of Fig.\ref{fig4} (dashed line) shows that this is approximately valid for the cos-pulse in the range $1<M<20$  .
\bibitem{ref34} Using again a genetic algorithm, we find that for our reference gain $M=7.4$ the fractional advancement of the pulse maximum would be $a_{m}/\tau_{p}=0.44$ instead of $0.32$ with the single-dip system and $0.29$ with our 2-pulse system.
\end{thebibliography}
\end{document}